\documentclass[authoryear,11pt]{elsarticle}

\usepackage[margin=0.80in]{geometry}
\usepackage{float}
\usepackage{adjustbox}
\usepackage{wrapfig}
\usepackage{soul}
\usepackage{footnote}
\usepackage{lineno}
\usepackage{xcolor}
\usepackage{multirow}
\usepackage{graphics}
\usepackage{graphicx}
\usepackage[font=small,skip=0pt]{caption}
\usepackage{subcaption}

\usepackage{amssymb,amstext,amsmath}
\usepackage{float}
\usepackage{booktabs}
\usepackage{array,ragged2e}
\usepackage{multirow}
\usepackage{algorithm}
\usepackage{algpseudocode}
\usepackage{pifont}
\journal{}

\begin{document}

\begin{frontmatter}



\title{Queue Length Estimation at Traffic Signals: Connected Vehicles with Range Measurement Sensors\vspace{-3mm}}
%\author[label1,label2]{Gurcan Comert, Anton Bezuglov}
%\address[label1]{Department of Physics and Engineering, Benedict College, 1600 Harden St., Columbia, SC 29204 USA}
%\address[label2]{Department of Math and Computer Science, Benedict College, 1600 Harden St., Columbia, SC 29204 USA}
\author[rvt]{Gurcan Comert\corref{cor1}}
\ead{gurcan.comert@benedict.edu}
\author[focal]{Mecit Cetin}
\ead{mcetin@odu.edu}
%\author[rvt]{Negash Begashaw}
%\ead{negash.begashaw@benedict.edu}
\cortext[cor1]{Corresponding author}
%\fntext[fn1]{This is the specimen author footnote.}
%\fntext[fn2]{Another author footnote, but a little more longer.}
\address[rvt]{Department of Computer Sc., Phys., and Eng., Benedict College, 1600 Harden St., Columbia, SC 29204 USA}
\address[focal]{Department of Civil and Env. Engineering, Old Dominion University, 135 Kaufman Hall, Norfolk, VA 23529 USA
\vspace{-10mm}}
%\author{Gurcan Comert and Anton Bezuglov}
%\address{Department of Physics and Engineering, Benedict College, 1600 Harden St., Columbia, SC 29204 USA\\
%Department of Math and Computer Science, Benedict College, 1600 Harden St., Columbia, SC 29204 USA}

\begin{abstract}
Today vehicles are becoming a rich source of data as they are equipped with localization or tracking and with wireless communications technologies. With the increasing interest in automated- or self- driving technologies, vehicles are also being equipped with range measuring sensors (e.g., LIDAR, stereo cameras, and ultrasonic) to detect other vehicles and objects in the surrounding environment. It is possible to envision that such vehicles could share their data with the transportation infrastructure elements (e.g., a traffic signal controller) to enable different mobility and safety applications. Data from these connected vehicles could then be used to estimate the system state in real-time. This paper develops queue length estimators from connected vehicles equipped with range measurement sensors. Simple plug-and-play models are developed for queue length estimations without needing ground truth queue lengths by extending the previous formulations. The proposed method is simple to implement and can be adopted to cyclic queues at traffic signals with known phase lengths. The derived models are evaluated with data from microscopic traffic simulations. From numerical experiments, the QLE model with range sensors improves the errors as much as $25\%$ in variance-to-mean ratio and $5\%$ in coefficient of variation at low $\leq20\%$ market penetration rates.

\end{abstract}


\end{frontmatter}

%\squeezeup
%\vspace{-3mm}
\section{Introduction}
\vspace{-3mm}
Tracking individual vehicles traversing the transportation network in real time has gained significant momentum lately due to recent developments in Connected Vehicles (CVs) technologies. Real-time data from CV enable observing complex system dynamics and support new applications to improve system efficiency and safety. However, not all vehicles may be equipped with CV technologies nor all drivers are willing to share their locations. Consequently, predicting the overall system behavior based on data from a fraction of vehicles has been an area of active research for more than three decades now.  
Queue Length (QL) is a fundamental performance criteria for any system with multiple users demanding service from a server as for real-time control of signalized intersections or port terminal gates. It is critical for applications such as emergency management and reliability of the network. Models based on performance measures can be input for traffic control, facility design, and construction planning purposes. Specifically, the real-time (adaptive) control demands high accuracy from such models. Today, vehicles are becoming a rich data and computational resource as they are equipped with localization or tracking, wireless communication technologies, and microprocessors. Furthermore, with the increasing interest on automated- or self- driving technologies, vehicles are also being equipped with range measuring sensors (e.g., LIDAR, stereo cameras, and ultrasonic) to detect other vehicles and objects in the surrounding environment. It is possible to envision that such vehicles could share their data with the transportation infrastructure elements (e.g., a signal controller) to enable various traffic mobility and safety applications. Data from these CVs could then be used to estimate the system state in real-time. Consequently, this paper is focused on estimation of QL from such CVs that may represent partial vehicles in the traffic or other systems (\cite{nam2009estimation}). Connected and autonomous vehicles (CAVs) or autonomous vehicles can also be treated similarly.
\vspace{-3mm} 
\subsection{Background}
Queue length estimation (QLE) from connected vehicle at traffic signals is a well-investigated problem. In basic form, there is no feedback from environment, thus, true queue length values are not known. This makes critical difference with the prediction formulations with series of true queue lengths are available where researchers used filtering, fit models, and deep learning methods (\cite{tiaprasertqueue,yin2018kalman,gao2020integrated,comert2019grey}). 

In the literature, researchers obtained CV based queues via reconstructing trajectories (\cite{argote2011estimation, cetin2012estimating, cheng2012exploratory, xu2017queue, rompis2018probe,zhao2019various}). Stochastic formulations are desirable as increase in number of related paper show. Researchers developed important models using similar approach. Within the scope and space, for more information on general queue length estimation, readers are referred to recent survey studies on QLE methods from CAVs (\cite{zhao2019various, guo2019urban}) and queue estimation bibliography by (\cite{asanjarani2017parameter}). Hence here, we focus on related previous works on stochastic QLEs from CVs are discussed. 

Stochastic models of QLEs from CVs mainly concerns with stationary (behavior after sufficiently long time interval) or non-stationary (short-time intervals, by cycle, and red phase). Such classification is impacted by treatment of arrival rate, distribution, and market penetration rate (MPR) (probe percentage). For instance, cycle-to-cycle estimation of QLs including green arrivals was modeled in (\cite{tan2019cycle}) and real-time applications with high frequency connected vehicle data was presented by(\cite{mei2019bayesian}). 

In queueing, detection the stopped vehicles accurately is crucial as it will be impacting the back of queue. This problem was addressed in (\cite{liu2019real}). Reported error levels in similar QLE studies may exceed $10 \%$ of true queue lengths (\cite{argote2011estimation}) which would require $80 \%$ CV penetration rate. In order to improve the accuracy, data fusion from different traffic surveillance technologies are also considered such as loops (\cite{li2013estimating,comert2013effect}), license plate recognition (\cite{tan2020fuzing}), and inclusion of radio frequency identification \cite{wu2013real} which acts as filtered QLE would give higher errors. Queue lengths for constant arrival rate at the end of red phase act like autoregressive (2) time series with large coefficients with opposite signs or moving average (1) with positive high coefficient. Therefore, filtering for a true QLs would over-smooth the series. With back of queue estimation and convex optimization, similar error levels at $p=20 \%$ was observed (\cite{yang2018queue}). Range sensor applications in traffic in some capabilities are discussed in (\cite{mandal2020measuring}). Authors, placed ultrasonic sensors capable Arduinos on lane markings and reported distance that vehicles could be detected for queue lengths. Although we seek higher penetration levels, efficient estimators performing at low MPRs would be desirable as we will need to address efficient data management. Question of how much data we need is important not to overload systems or security reasons. Partially supporting these, in an interesting report, researchers found in a simulation environment, over 200 vehicles at $100 \%$ penetration level signal control from CVs becomes difficult (\cite{university2016multi}) due to computational load.    

%\textcolor{red}{Transition to specific problem:}
In this study, simple plug-and-play expressions are developed for QLE from CVs in the case of connected vehicles are also equipped with a range sensor by extending the authors' previous formulations in (\cite{comert2009queue,Comert2011,comert2013simple,comert2016queue}). Derived formulate in this paper incorporate the measured range from each CV as an additional source of real-time data supplementing the typical CV information (e.g., location and time). In a standing queue, the range will inform whether there is a vehicle behind a CV. Steady-state behavior of estimated QLs and their errors are formulated. These are then compared against and validated with results from point queue as well as microscopic simulations.
%In an interesting continuation work of \cite{Leeuwaarden2006}, \cite{timmerman2017fixed} provides.
\vspace{-3mm}   
\subsection{Contributions of this study}
Our modular expectation-based approach enables application of filtering algorithms to deal with measurement or low MPR related errors. Note also that the estimation problem is interesting where only assume partially observed system. Thus, for any similar problem such as autonomous vehicles, our approach can be adopted. One of limitations for connected vehicles with low MPRs, as we discussed in (\cite{comert2016queue}), is distinguishing whether the last connected vehicle is the last vehicle in the queue. This information would be available from a range sensor (ultrasonic , LIDAR) \cite{cetin2017exploring} showed the accuracy levels in LIDAR-based estimations. Moreover, presented formulae do not depend on any primary parameter such as arrival rate or MPR. So, method can be used for time dependent arrival rates or $p$. Estimation of these primary parameters are becoming interest also (\cite{comert2016queue,zhao2019estimation,van2018estimation}).
  
Impact of additional sensors in terms of estimation errors for different arrival rates $\lambda$ and MPR (probe percentage) $p$ are presented. This research gives insights about the benefit of additional data from range measuring sensors that can be common with the upcoming autonomous vehicles. The results can be incorporated for more accurate queue lengths and delay calculations in new control heuristics (\cite{comert2009incorporating}).
\vspace{-5mm}
\section{Problem Statement and Notation}
\label{sctprb}
\begin{figure}[!htb]
\centering
\includegraphics[scale=4.5]{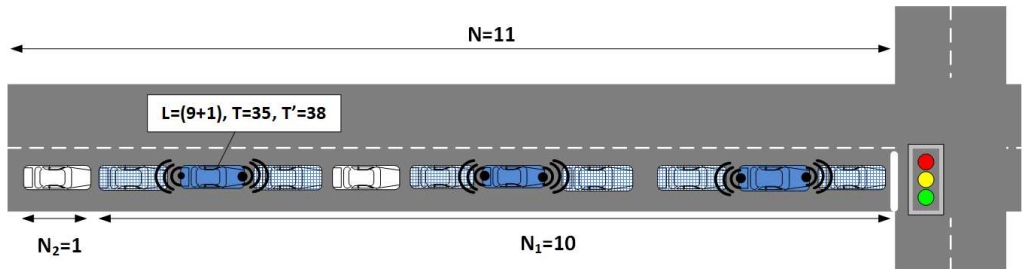}
\caption{Approach lane right before the green phase}
\label{fig_int}       
\end{figure}

Fig.~\ref{fig_int} shows a view of an approach lane right before turning green, vehicles with darker colors represent CVs and shaded ones depict the detected vehicles via range sensors. Location of the connected vehicle information in queue $L$ are inferred as number of vehicles. Number of CVs ($M$) is 3, location of the last CV ($L$) is $9$, and queue time tag of the last probe $T$ is $35$ seconds (s). Moreover, last CV is able to detect the vehicle behind it, so, $L'$=($9+1$) and queuing time of the following vehicle detected by sensor on the last CV is $T'=38$ s. 

Basic problem is defined as deriving estimators for the total queue length $N$ given the information from CVs and other detected vehicles by CVs' range measuring sensors such as LIDARs (i.e., CV locations and time tags). Range measuring sensors provide the critical information of whether the last CV actually is the last vehicle in the queue (\emph{i.e.}, there is at least one more vehicle joined the queue after it). The queue length ($N$) is 11 which is estimated in real-time at the end of red phase $R$ using the location and the time of CVs ($L'=10$, $T'=38$, $M=3$, and $R$). Estimators are developed for known $R$, but they can be used for actuated or adaptive signal phase lengths at the time instance of interest.
\vspace{-10mm}
\subsection{Limitations of the study} 
In this study, we aim to develop analytical input-output models. For simplifications we assume following. 
\begin{enumerate}[(i)]
\item Approach lane has no capacity limit and there is no loss time due to acceleration. These assumptions are for adopted overflow queue models (\cite{Haight1959,Ohno1978,Medhi1991,Heidemann1994}). Usefulness of the models in the proposed estimators are shown via comparing with errors from microscopic traffic simulations (i.e., VISSIM) which provide more realistic arrivals and queuing at the signal. 
\item Poisson arrivals to the approach lane are assumed which can describe isolated intersections, low demand scenarios of more complex intersections, and enable simple closed-form derivations.  
\item Lane level accuracy for CV locations is assumed. Position and time tags are accurately received. Connected vehicle architectures especially at intersections will satisfy this by map information and timing passed by roadside units. When location is provided as latitude and longitude, using average vehicle length or passenger car equivalencies, spot in number of vehicles can be estimated. Basic safety messages also include vehicle type, using vehicle lengths by type locations can also be estimated.  
\item Uniform vehicle type is assumed. Mixed traffic could be incorporated to the models by arrival rates in Poisson arrivals (\cite{walraevens2017delay}). Impact of different vehicle types would impact intersection capacity and location information.
\item Connected vehicles are able to sense accurately if there is a vehicle behind with the help of range sensors.  
\end{enumerate}

Estimators presented are unbiased under these assumptions. They would hold for estimating real-time queues at isolated signals, signals with low to moderate volume-to-capacity ratios, Poisson, or wherever approximately random arrivals can be assumed (\cite{Newell1982,fiems2019taylor}). Such QLEs are going to be unbiased. In fact, any approach queue length can be estimated given formulae. Amount of deviations from random arrivals will be reflected into the estimation errors as bias.
\vspace{-5mm}
\section{Analysis for the Case without Overflow Queue}
\label{sctwoq}
\vspace{-2mm}
First, it is assumed that vehicles that have arrived in one given cycle clear the intersection in that cycle (the overflow queue ($Q$) is zero). At each cycle, the total number of vehicles in the queue $N$ is estimated given $L$, $T$, and primary parameters arrival rate $\lambda$, $p$, and $R$. Throughout the examples, signal related parameters are chosen to be effective red duration of $45$ $s$, green time of $43.2$ $s$, and discharge rate $1.8$ seconds per vehicle ($spv$). Other selected parameter values are $\lambda=\{0.111, 0.133, 0.163, 0.190, 0.218, 0.239, 0.267\}$ $vps$ which correspond to volume-to-capacity ratios of $\rho=\{0.41, 0.49, 0.60, 0.70, 0.80, 0.88, 0.98\}$ and $0.001 \leq p \leq 1.00 $ (\cite{comert2016queue}).

As shown in Eq.~\ref{eqn_ewoq}, estimator of the queue length $N$ can be written as summation of expected values of the number of vehicles known until last CV $N_1$ and inferred $N_2$ after this CV for given conditions: (i) whether the last CV is also the last vehicle $I(L=l_v)=1$ and (ii) the last CV is not the last vehicle $I(L\neq l_v)=1$. Here, $I()$ is the indicator function. 
\begin{eqnarray}
E(N|{L=l,T=t})= \begin{cases} [E(N_1|{L=l,T=t})+E(N_2|{L=l,T=t})]I(L=l_v)+  \\
[E(N_1|{L=l,T'=t'})+E(N_2|{L=l,T'=t'})]I(L\neq l_v) \end{cases} 
\label{eqn_ewoq}  
\end{eqnarray}

In Eq.~(\ref{eqn_ewoq}), for condition (i) $I(L=l_v)$, $N_1$ is known and $N_1=l$ and $E(N_2|{L=l,T=t)}$ is number of arrivals after the last CV during $(R-T)$=$\delta$ which is zero leading to estimator with no error $N=l$. For condition (ii) $(L\neq l_v)$, after the last CV, only non-CVs would arrive, thus $N_1=l+1$ and $\delta=(R-T')$. Arrival rate during $\delta$ is $(1-p)\lambda$. Denoting $(1-p)\lambda$ by $\theta$, we can simply write $N_2|{L=l,T=t'}$$\sim$Poisson($\theta\delta$). Simplified estimator is given in Eq.~(\ref{eqn_ewoq2}). 

For the estimation error when first condition, queue length in Eq.~(\ref{eqn_ewoq2}) is known ($N=l$), thus, the error is $zero$. For the second condition, first term is known ($N_1=l+1$) and there is no error. The second term is $\sim$Poisson($\theta\delta$) and the variance is $\theta\delta$. Thus, the conditional variance of Eq.~(\ref{eqn_ewoq}) for $L\geq0, T\geq0$ can simply be written as Eq.~\ref{eqn_vwoq}. Practically, given the time information and if there is a vehicle behind or not, we can calculate the estimation errors under the assumptions listed in \ref{sctprb}.
%\vspace{-3mm}
\begin{eqnarray}
E(N|{L=l,T=t})=\begin{cases} l, & L=l_v\\ l+1+\theta\delta, & \delta=(R-t'), L\neq l_v \end{cases} 
\label{eqn_ewoq2}  
\end{eqnarray}
\begin{equation}
V(N|{L=l,T=t})=\theta(R-t')I(L\neq l_v), L\geq0,\ T\geq0
\label{eqn_vwoq}  
\end{equation}

Formula for the errors can be expressed as deviations from $D=N-E(N|L=l,T=t)$. Thus, the variance of $D$ for $L\geq0$ can be written as $V(D)=E[V(N|L=l,T=t)]$ and given in Eq.~(\ref{eqn_vdwoq}). Suppose that $E(T)$ and $E(T')$ are known (without and with range sensor, respectively), errors can be evaluated and simplified to Eq.~\ref{eqn_vdwoq}.
%\vspace{-1mm}

\begin{eqnarray}
V(D)=E[\theta\delta]=\theta[R-E(T')]P(L\neq l_v)=\frac{p (1-e^{-p\lambda R(1+p)})}{(1+p)},\ t<R	
\label{eqn_vdwoq}  
\end{eqnarray}
where, $P(L\neq l_v)=E(1-e^{-\lambda(R-t)}),\ t<R$ and $P(L=l_v)=1-P(L\neq l_v)$ can be interpreted as $L=l_v$ probability of last CV is the last vehicles meaning no arrival during $\delta=(R-t)$ (i.e., $e^{-\lambda(R-t)}$). Using distribution of $T$ (\cite{comert2013simple}) $f(t)=\frac{\lambda p e^{-p\lambda(R-t)}}{(1-e^{-\lambda pR})},\ t<R$ is input for scenario probabilities $P(L\neq l_v)$ as well as for the expected value of $T'$ respectively. They can then be incorporated to compute the variance given in Eq.~(\ref{eqn_vdwoq}).
%\vspace{-3mm} 
\subsection{Derivation of the expectations}
\label{sctdists}
Revising the the probability distributions derived in \cite{comert2013simple}, queue length estimators and the errors are derived. Corresponding $E(L)$ and $E(L')$ revised for range sensor incorporation which are given in Eq.~(\ref{eqn_el}). The impact of sensor information is simply in the denominator adding one non-CV $(1-p)$.
\begin{eqnarray}                                                                
E(L)=\lambda R-[(1-e^{-\lambda pR})/p],\ m>0 \nonumber \\
E(L')=\lambda R-[(1-e^{-\lambda pR})/(p+(1-p))],\ m>0   	 		   			\label{eqn_el}  
\end{eqnarray}
%\vspace{-3mm}

\begin{figure}[!h]%[!htb]
\centering
\includegraphics[scale=.80]{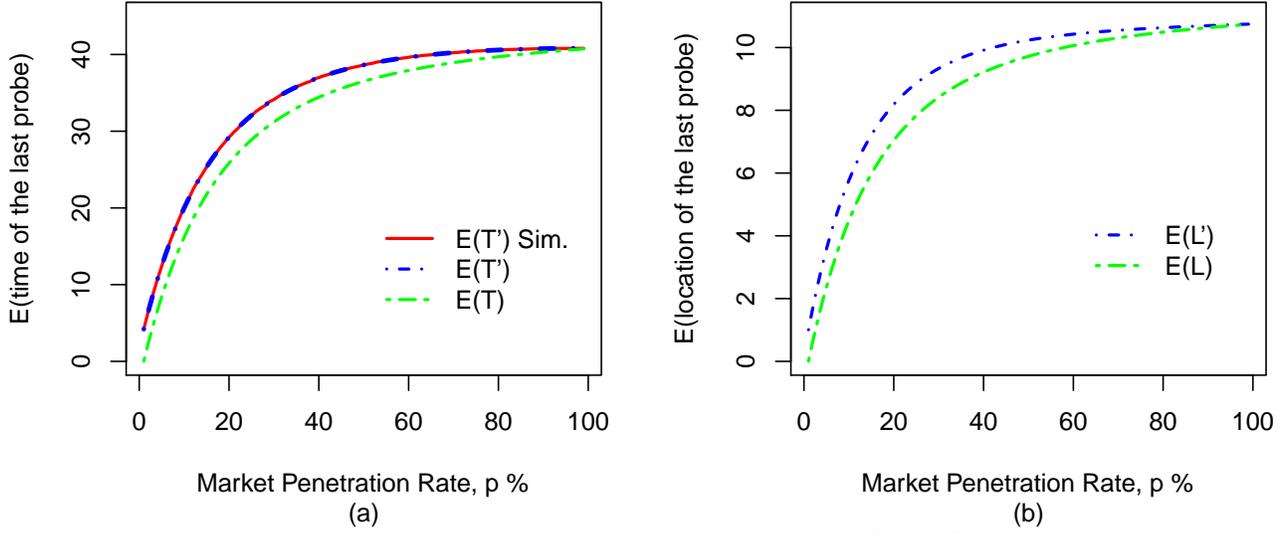}
\caption{Impact of range sensor on $\{E(T),E(L)\}$ shown as $\{E(T'),E(L')\}$ for $\lambda=0.239$ $vps$}
\label{fig_el}       
\end{figure}

Similarly, in Eq.~(\ref{eqn_et}), $E(T)$ and $E(T')$ are provided. Impact is simple again by adding the term $(1-p)\lambda$. The revised variance of $D$ ($V(D)$) can be calculated using Eq.~(\ref{eqn_et}) in Eq.~(\ref{eqn_vdwoq}) and shown in Eq.~(\ref{eqn_crrlry}). Note that the QLE error formula without sensors is $V(D)=E[\theta\delta]=\theta[R-E(T)]$=$(1-p)(1-e^{-p\lambda R})/p$.   

\begin{eqnarray}                                                            E(T)=R-[(1-e^{-\lambda pR})/\lambda p],\ m>0 \nonumber \\
E(T')=R-[(1-e^{-\lambda pR})/\lambda p+(1-p)/\lambda],\ m>0   	 		   	
\label{eqn_et}  
\end{eqnarray}
%\vspace{-1mm}
One of the main contribution of \cite{comert2013simple} was to be able to compute $E(L)$ and $E(T)$ from one another, thus, $E(T')$ can be utilized to calculate $E(L')$ and shows the impact of range sensor in Fig.~(\ref{fig_el}) which are derived using Eqs.~(\ref{eqn_et}) and (\ref{eqn_el}). This error (Eq.~\ref{eqn_crrlry}) can be calculated using estimators for $p$ and $\lambda$ and signal timing information. 
%\vspace{-4mm}

\begin{eqnarray}   
V(D)=E[\theta\delta]=\theta[R-E(T')]P(L\neq lv) \nonumber \\
=(1-p)\lambda[R-(R-(1-e^{-p\lambda R})/\lambda p+(1-p)/\lambda)]P(L\neq lv) \nonumber\\
=[1-p][(1-e^{-p\lambda R})/p-(1-p)]P(L\neq lv)
\label{eqn_crrlry} 
\end{eqnarray}
where, $P(L\neq lv)=[1-p(1-e^{-(1+p)\lambda R})/(1+p)]$.
%\squeezeup
%\vspace{-1mm}

To verify the formulation above, the results are compared with simple point queue simulations with Poisson arrivals coded in \textsf{R}. From Fig.~\ref{fig_el}-a, it can be seen that $E(T')$ is following the simulated values very closely. As $p$ is approaching $100\%$, the $E(T')$ is getting closer to the red duration (e.g., $45$ seconds). In addition, Fig.~\ref{fig_el}-b shows the values of $V(D)$s computed with the analytical models with and without range measuring sensors. These are then compared against simulations with $100,000$ replications. As it is evident analytical results follow simulation results very closely. 
%\squeezeup
\begin{figure}[h]%[!htb]
  %\vspace*{1mm}
\centering
\includegraphics[scale=.80]{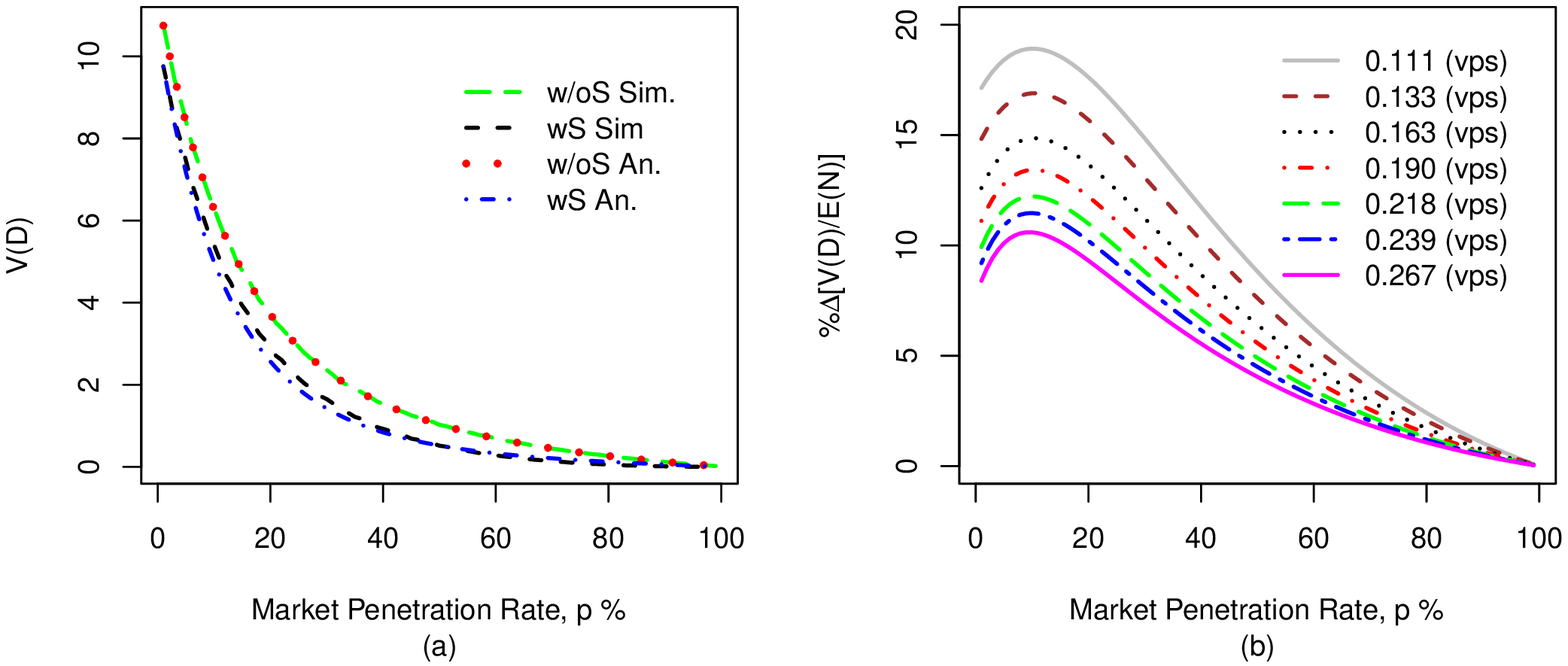}
\caption{Improvements in QLE errors when each CV also provides its following vehicle's position in the queue  (various $\lambda$ levels)}
\label{fig_errorswoQ}       
\end{figure}

Figs.~\ref{fig_errorswoQ} a-b are given to illustrate how much improvement is achieved when CVs are also equipped with range measuring sensor, i.e., these graphs show the additional reduction in QLE error when $N$ is estimated from CVs that also provide information about their following vehicles (as compared to CVs that provide only their own positions). In Fig.~\ref{fig_errorswoQ} a, this improvement is shown via graphing percent differences of $V(D)/E(N)$s (i.e., variance-to-mean ratio or dispersion index) with respect to probe percentages $p$. Figure depicts maximum gain (e.g., close to $20 \%$ at $\lambda=0.133$ $vps$) at about $p=10 \%$ and increasing as $\lambda$ decreases. Figure suggests that about $20\%$ probe vehicles are need to experience the maximum impact ($\approx 10\%$). Lower $p$ values are still showing $8$ to $10\%$ improvements relative to mean QLs. And, the impact is diminishing as $p$ increases where $P(L=lv)$ also get higher.  
%\squeezeup
\begin{figure}[!h]%[!htb]
\centering
\includegraphics[scale=.80]{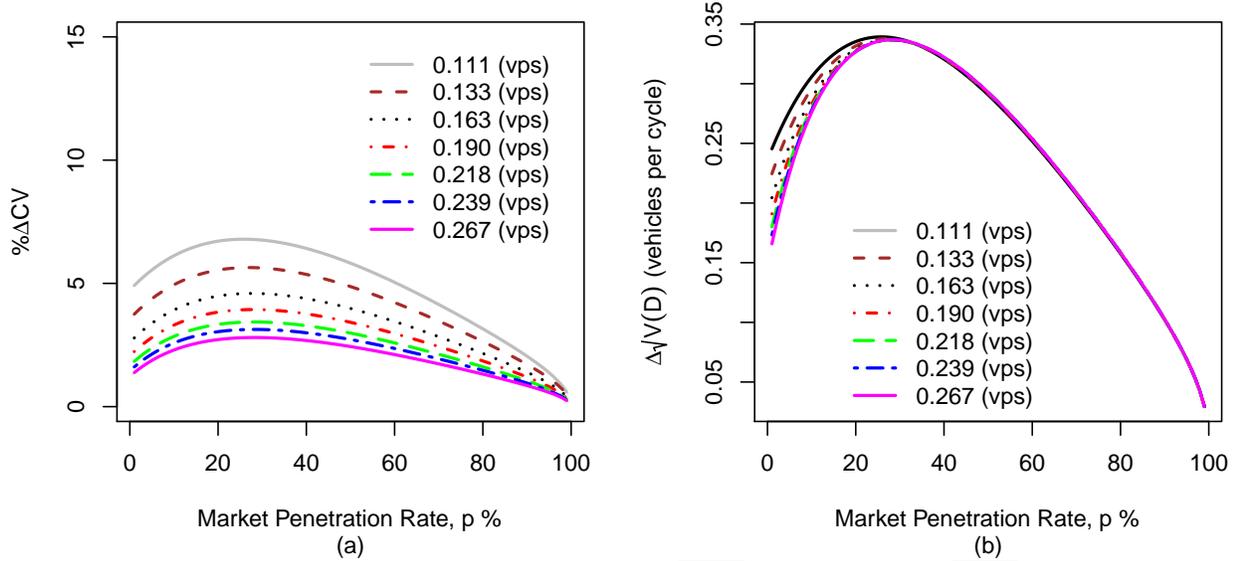}
\caption{Relative errors $\% \Delta CV=\sqrt{V(D)}/E(N)$ and $\% \Delta \sqrt{V(D)}$}
\label{fig_cvwoQ}       
\end{figure}

Figs.~\ref{fig_cvwoQ} a-b demonstrate the impact of range sensors in terms of $\%$ difference of $\sqrt{V(D)}/E(N)$s (i.e., coefficient of variations) and difference of $\sqrt{V(D)}$s changing as functions of market penetration rate $p$ at different $\lambda$ values. The maximum improvement (e.g., close to $7 \%$ for $\lambda=0.133$ $vps$) at about $p=30 \%$ and increasing as $\lambda$ decreases. Similarly, estimation error differences are higher for lower $\lambda$ values. They get very close to each other after about $30 \%$ $p$ level. Difference of $\sqrt{V(D)}$ values are given in Fig.~\ref{fig_cvwoQ} b that are QLE error differences in terms of number of vehicles. Again, maximum impact $0.35$ vehicle per cycle is observed at $p \approx 30\%$. This behavior would make such estimator more suitable for intersections with low demand or rural intersections. Desirably, this would also satisfy the assumptions for the derivations. 
%\vspace{-80pt}
%\vspace{-20pt}
%\vspace{-7mm}
\section{Analysis for the Case with Overflow Queue}
\label{sctwq}
%\vspace{-2mm}
When the overflow queue from signal cycles is incorporated in the queue length estimation, we can express queue length $N$ as the sum of the overflow queue $Q$ and the new red duration arrivals $A$. Then, for any cycle, based on the last CV information (e.g., CV in $Q$ or $A$) and whether it is the last vehicle ($L=lv$) (i.e., detected by range sensor), we can model the estimator with the help of five possible scenarios:
%\vspace{-3mm} 
\begin{enumerate}[(i)]
	\item The last CV is within new arrivals and it is the last vehicle, $I(l\in A \land L=lv)$.
	\item The last CV is within new arrivals and it is not the last vehicle, $I(l\in A \land L \neq lv)$.
	\item The last CV is in the overflow queue and it is the last vehicle, $I(l\in Q \land L=lv)$.
	\item The last probe may be within the overflow queue and it is not the last vehicle, $I(l\in Q \land L\neq lv)$. 
	\item There is no CV  in the queue, $I(l=0)$.  
\end{enumerate}
	
\begin{figure}[!h]%[!htb]
\centering
\includegraphics[scale=.80]{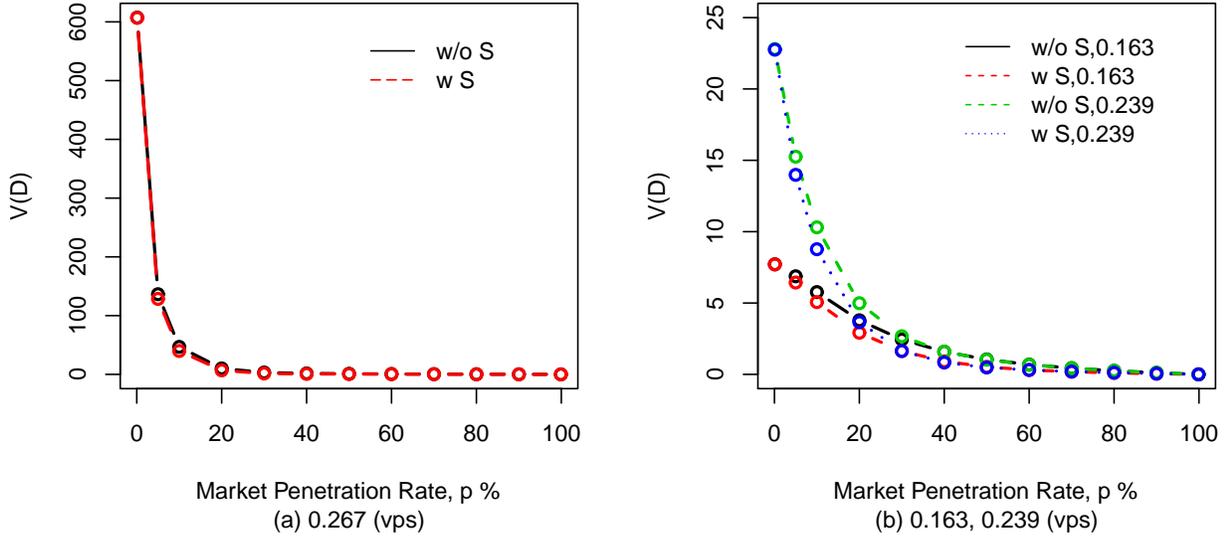}
\caption{Estimation errors from VISSIM with and without $Q$ results for different $\lambda$ and $p$  values}
\label{fig_vard}       
\end{figure}

Considering these scenarios, we can write $N|L,T$=$I(L=lv)[Q|L,T + A|L,T]+I(L \neq lv)[Q|L,T + A|L,T]$. Then, the estimator for the total queue length given CV information can be written as Eq.~(\ref{eqn_ewq}). When utilized, Fig.~\ref{fig_vard} was given for the average estimation errors $V(D)$s from with and without range sensors. From the figure, similar to $Q=0$ case, improvements with sensors are higher in between $p$=$0.05-30\%$. The estimator with overflow queue also contains a revised queue joining time $\tau$ that defined as the remaining last CV's joining time from previous green duration. These observations are desired and expected. Benefit at lower CV penetration levels would be more cost saving. As $p$ gets higher, it is expected that more CVs will be the last vehicle in the queue, thus, saving would be less from the range sensor fusion.    
\begin{equation}   
E(N|L=l,L=lv,T=t)=\begin{cases} I(l\in Q \land L=lv)[l+\theta R]+\\ I(l\in Q \land L \neq lv)[l+1+\theta (C-\tau')+\theta R]+\\ I(l\in A \land L=lv)[l]+\\ I(l\in A \land L \neq lv)[l+1+\theta (R-t')]+\\I(l=0)[(1-p)(E(Q)+\theta R)].\end{cases}   
\label{eqn_ewq}  
\end{equation}
%\vspace{-10pt}
%\squeezeup                 

Average overflow queue in Eq.~(\ref{eqn_ewq}), $E(Q)$=$\frac{\rho^2}{2(1-\rho)}$, is adopted from (\cite{Akcelik1980}) where $\rho=\lambda C/X$ with $X=24$ vehicles per cycle in our examples, $\lambda$ arrival rate per second, and $C$ is cycle time $90$ seconds. Alternatively, from~(\cite{Viti2006,Medhi1991}) different average overflow queue formula can be adopted $E(Q)$=$1.5(\rho-\rho_{o})/(1-\rho)$, $E(Q)$=$\frac{(2\rho-1)\rho^4}{2(1-\rho)}$, and $E(Q)$=$[\frac{\rho}{2(1-\rho)}]e^{-[\frac{(1-\rho)\sqrt{R/2}-R(1-\rho)^2}{4}]}$ where $\rho_{o}$=$0.67+X/600$. The queue length on the approach lane is assumed to be \textit{zero} for all examples in the study. Using the scenario probabilities in Eq.~(\ref{eqn_en}), we can obtain Eq.~(\ref{eqn_ewq}).
%\begin{equation}   
%E(N|S=s,T=t)=\begin{cases} I(l\in Q)[l+\theta (C-t')+\theta R]+\\ I(l\in A)[l+\theta\delta]+\\I(l=0)[(1-p)(E(Q)+\theta R)]\end{cases}   
%\label{eqn_ewq}  
%\end{equation}
%\begin{equation}   
%P(L=l)=\begin{cases}[1-\sum_{q=0}{(1-p)^q P(Q=q)}] e^{-\lambda Rp},l\in %Q\\1-\sum_{a=0}{(1-p)^{a}P(A=a)}, l\in A\\ \sum_{q=0}{(1-p)^q P(Q=q)e^{-\lambda %Rp}}, l=0\end{cases} 
%\label{eqn_sp}  
%\end{equation}
\begin{equation}   
E(N)=\begin{cases} P(l\in Q \land L \neq lv)[E(L)+\theta (C-E(\tau'))+\theta R]+\\ P(l\in Q \land L=lv)[E(L)+\theta R]+\\ P(l\in A \land L \neq lv)[E(L)+1+\theta(R-\tau)]+\\ P(l\in A \land L=lv)[E(L)]+\\ P(l=0)[(1-p)(E(Q)+\theta R)]\end{cases}   
\label{eqn_en}  
\end{equation}

Average or steady-state estimation error, $V(D)$s, can be given as in Eq.~(\ref{eqn_vdwq}) with $V(Q)$ in Eq.~(\ref{eqn_vq}) from~(\cite{Medhi1991}).
\begin{equation}   
V(D)=\begin{cases} P(l\in Q \land L \neq lv)[\theta (C-E(T'))+\theta R]+\\ P(l\in Q \land L=lv)[\theta R]+\\ P(l\in A \land L \neq Lv)[(1-p)(1-e^{-(1+p)\lambda R})/(1+p)]+\\P(l=0)[(1-p)(V(Q)+\theta R)]\end{cases}   
\label{eqn_vdwq}  
\end{equation}

\begin{figure}[!h]%[!htb]
\centering
\includegraphics[scale=.80]{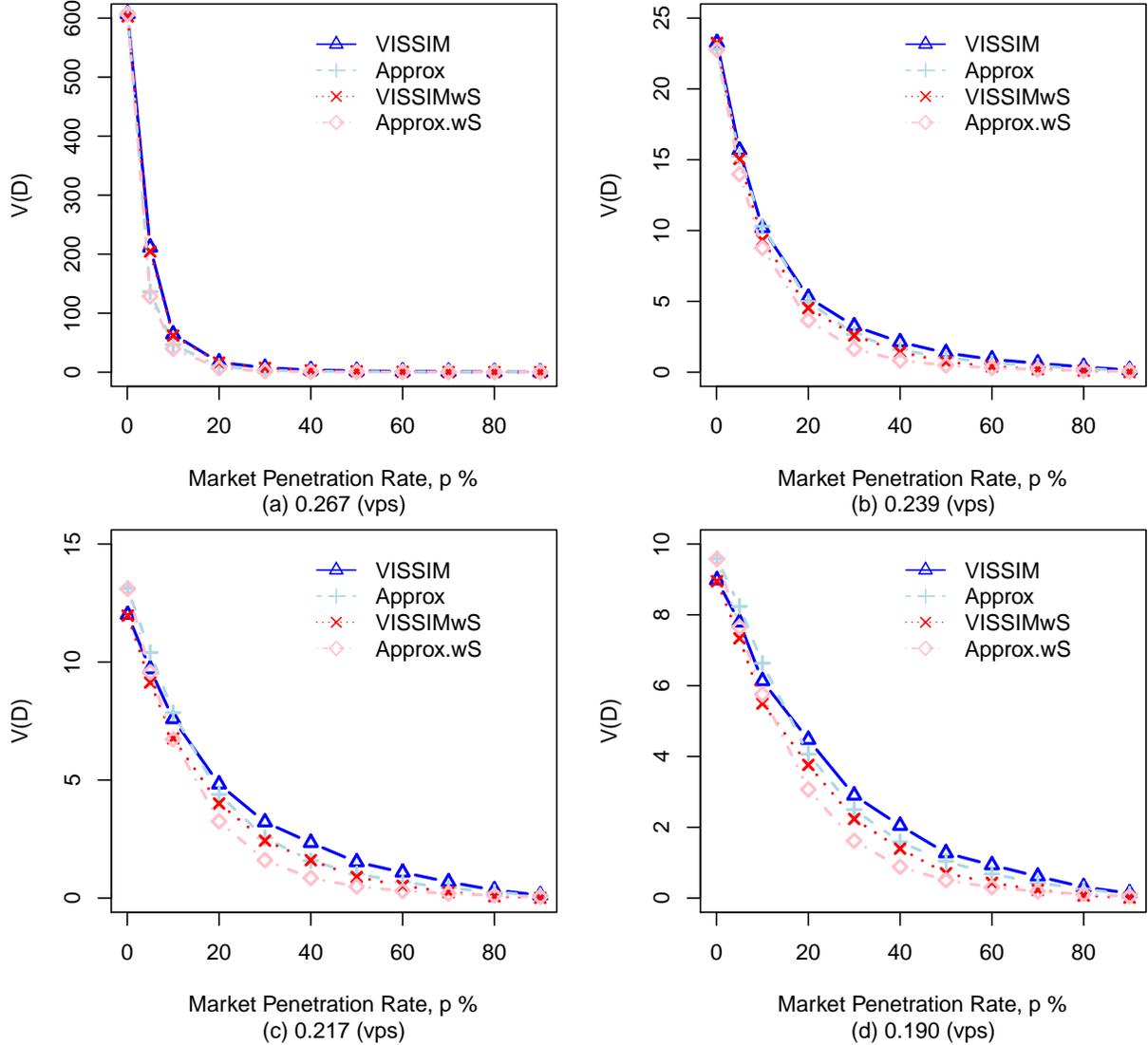}
\caption{Comparing estimation errors obtained from VISSIM and approximations}
\label{fig_comp}       
\end{figure}

Probability distributions of $Q$ and $\tau$ are needed to calculate Eq.~(\ref{eqn_vdwq}). These distributions can be obtained numerically (\cite{Olszewski1990,Viti2010}) or by solving the probability generating function (\cite{gray1992m, Tarko1994,Mung1996,Broek2006}). But, no exact simple form for $V(D)$ is available. Thus, in Section~\ref{sctapprox} we used approximations in the estimators and the errors and evaluated them with microsimulation results. Figs.~\ref{fig_comp} a-b are derived using the approximation formulae and microsimulations using Eq.~(\ref{eqn_vdwqapprox}). Details of the approximate estimator is explained next by utilizing steady-state $V(Q)$s by the formulae adopted from (\cite{Medhi1991,Fu2000}). The approximate $V(D)$s are also given against VISSIM results in Figs.~\ref{fig_vard}-\ref{fig_cvs}. Relative errors are expressed in coefficient of variations (i.e., $100\sqrt{V(D)}/E(N)$ or $\%\Delta CV$). Figures also given as $\%\Delta \sqrt{V(D)}$ from the two evaluations.  
%\squeezeup
%\vspace{-5mm}

\section{Approximate Inference for $E(N)$ and $V(D)$}
\label{sctapprox}
%\vspace{-2mm}
Approximation formulas for the estimators can be derived simply quantifying the impact of overflow queue in terms of $E(Q)$ and $V(Q)$ in Eq.~(\ref{eqn_eq}) and Eq.~(\ref{eqn_vq}), respectively. Note that these formulas assume $M/G/1$ queue and they are based on the Pollaczek-Khintchine formula (\cite{Haight1959,Heidemann1994,Ohno1978}) and adopted from (\cite{Medhi1991}). Numerical values are determined and discussed in the simulation set up (\cite{comert2016queue}), mean discharge rate from the intersection is determined as $1.8$ seconds per vehicle with the capacity for effective green of 43.20 seconds ($(43.20/1.80)$=$24$ vehicles per cycle). Demand levels of $\{600, 700, 800, 900, 985\}$ vph that corresponds to volume-to-capacity ratios of $\rho$=$\{0.60, 0.70, 0.80, 0.88, 0.98\}$ are used in the numerical examples. VISSIM is run $1000$ cycles ($25$ hours) for $3$ random seeds for 11 $p=\{0.1\%,5\%,10\%...,90\%\}$ CV percentages at each demand level. Average of $3$ random seeds are reported for estimation errors. Default queue definition of VISSIM in terms of minimum speed$\leq10$ kilometers per hour is used.  For detailed discussions, readers are referred to (\cite{Akcelik1980,Medhi1991,Fu2000,Viti2006}).

Note that we are presenting the steady-state QLE errors incorporating overflow queue moments, however, cycle-to-cycle formulas were also given $E(Q_i)$ and results are given as figures in Section \ref{sctest}. $E(Q_i)$ and $V(Q_i)$ are adopted from  (\cite{Akcelik1980}) and (\cite{Viti2006}), respectively.    
%Similar expressions for cycle-to-cycle expectation and variance for the overflow queue can be found in (\cite{Viti2006}) as presented and evaluated in (\cite{comert2013simple}).}
%This section describes the approximation models for $E(N)$ and $V(D)$ with and without range sensor given in Eqs.~(\ref{eqn_en}), (\ref{eqn_vdwqapprox}), (\ref{eqn_vdwqapprox2}). 
%The models are obtained by utilizing $E(Q)$ and $V(Q)$. They depend only on probe proportion $p$, arrival rate $\lambda$, and the signal phase lengths. Unlike prediction methods, they do not require true queue lengths to be known. Under the assumption of undersaturated conditions, the overflow queue occurs from occasional random demand fluctuations in traffic. 
%Applicable in similar conditions, the Pollaczek-Khintchine (K-P) formula for the steady-state $M/G/1$ queue is found appropriate to be used in the estimation models (\cite{Haight1959,Heidemann1994,Ohno1978}). Adopted K-P formula assumes a constant queue discharge rate (e.g., $0.555$=$1/1.8$ vehicle per second) and a linear relation between overflow queue $Q$ and overflow delay $W$ (i.e., the Little's law $Q$=$\lambda W$) (\cite{Heidemann1994,Ohno1978}). Steady-state expectation and variance for the overflow queue are given in Eqs.~(\ref{eqn_eq}) and~(\ref{eqn_vq}). 
\begin{equation}
E(Q)=\frac{\rho^2}{2(1-\rho)}
\label{eqn_eq}
\end{equation}                 
\begin{equation}
V(Q)=[4(1-\rho)\rho^3+3\rho^4 ]/(12(1-\rho)^2)
\label{eqn_vq}
\end{equation}                 

Simply, queue length estimation approximations for expected value and errors are obtained by incorporating Eqs.~(\ref{eqn_eq}) and~(\ref{eqn_vq}) into appropriate cases in Eqs.~(\ref{eqn_ewq}) and (\ref{eqn_vdwq}). Resulting approximations with range sensors are then expressed as  Eqs.~(\ref{eqn_ewqapprox}) and~(\ref{eqn_vdwqapprox}). Scenario probabilities can be simply approximated by quadratic equations. 
\begin{equation}   
E(N)=\begin{cases} P(l\in Q)[E(Q)+\theta R]+\\ P(l\in A)[E(Q)+\lambda R]+\\I(l=0)[(1-p)(E(Q)+\theta R)]\end{cases}   
\label{eqn_ewqapprox}  
\end{equation}

\begin{equation}   
V(D)=\begin{cases} P(l\in Q)[(1-p)(1-e^{-pE(Q)})/p+\theta R]+\\ P(l\in A)[(1-p)(1-e^{-p\lambda R})/p]+\\P(l=0)[(1-p)(V(Q)+\theta R)]\end{cases}   
\label{eqn_vdwqapprox}  
\end{equation}
where, approximate probabilities are given in Eq.~\ref{eqn_spapprox}.
%\vspace{-2mm}
\begin{equation}   
P(L=l)=\begin{cases}e^{-p\lambda R}-e^{-p(\lambda R+(9.87p^2-4.62p+0.991)E(Q)},L\in Q \\ 1-e^{-p\lambda R}, L\in A\\ e^{-p(\lambda R+(9.87p^2-4.62p+0.991)E(Q))}, L=0\end{cases} 
\label{eqn_spapprox}  
\end{equation}
\begin{equation}   
V(D)=\begin{cases} P(l\in Q)[(1-p)p(1-e^{-(1+p)E(Q)})/(1+p)-p\theta R]+\\ P(l\in A)[(1-p)(1-e^{-p\lambda R})/p-(1-p)][1- p(1-e^{-(1+p)\lambda R})/(1+p)]+\\P(l=0)[(1-p)(V(Q)+\theta R)]\end{cases}   
\label{eqn_vdwqapprox2}  
\end{equation}

\begin{figure}[!h]%[!htb]
\centering
\includegraphics[scale=.80]{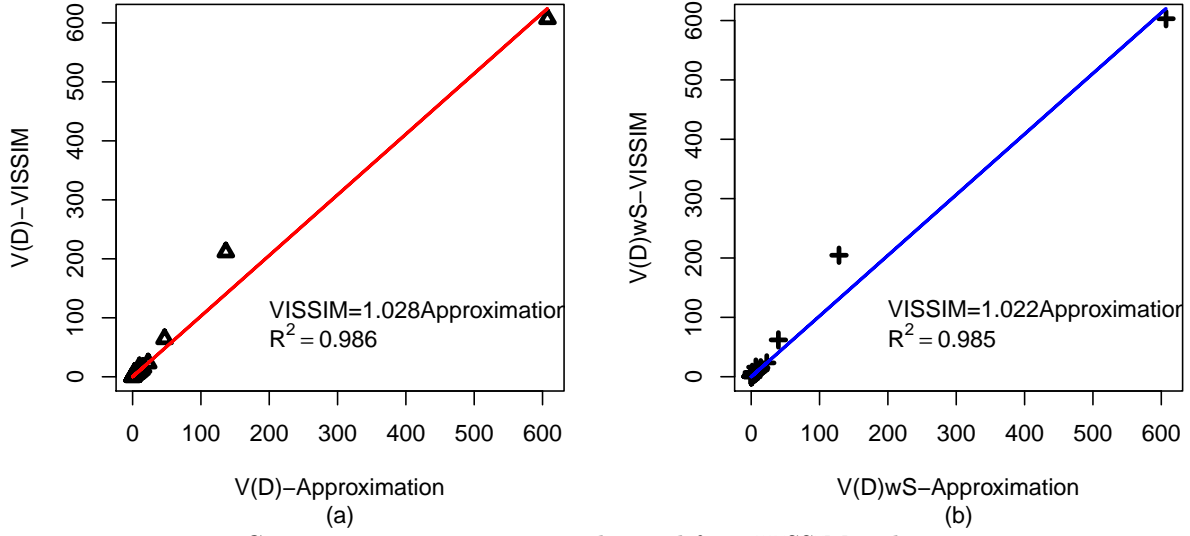}
\caption{Comparing estimation errors obtained from VISSIM and approximations}
\label{fig_reg}       
\end{figure}

Performance of the approximations are presented in Figs.~\ref{fig_comp} a-d. It can be seen that approximate errors are following VISSIM values closely for both with and without range sensors at all $\rho$ and $p$ levels. Moreover, Fig.~\ref{fig_reg} presents linear regression lines without intercepts for $V(D)$s from VISSIM simulations and approximate models without and with range sensor. Regression lines are given for estimated and simulation results. The lines show high correlations between analytical approximations and VISSIM errors without and with range sensor at $R^2$=$0.986$ and $R^2$=$0.985$, respectively. 
%\squeezeup

\begin{figure}[!h]%[!htb]
\centering
\includegraphics[scale=.80]{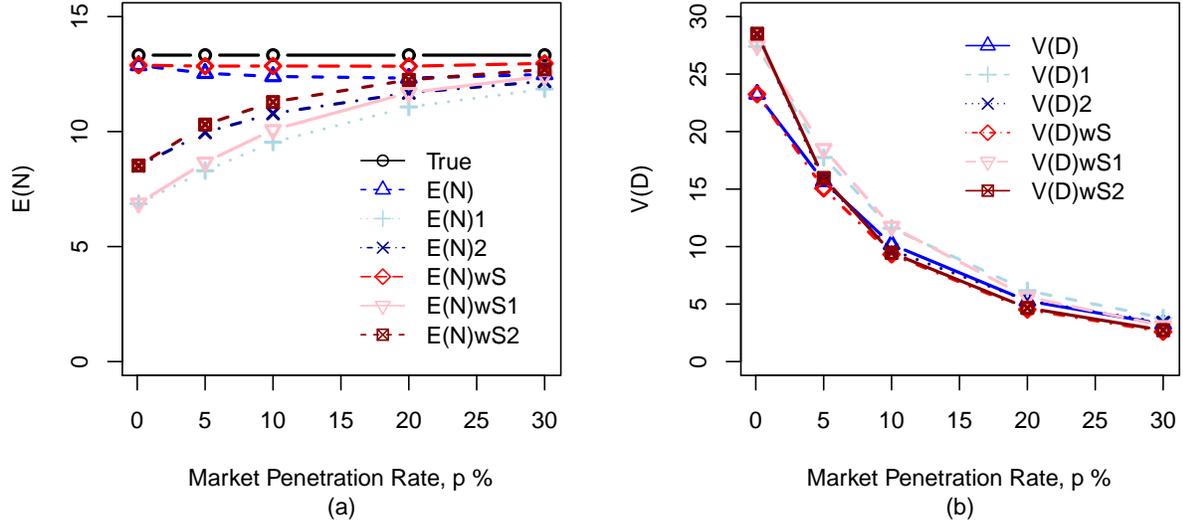}
\caption{Estimation errors with and without results for different $\lambda$ $vps$ and $ps$}
\label{fig_varden}       
\end{figure}

Fig.~\ref{fig_varden}-a shows Eq.~(\ref{eqn_vdwqapprox}) values along with VISSIM QLs for various $\rho$ levels. Developed approximation for QL is a simple formula that needs $\rho$, $\lambda$, and signal timing. Fig.~\ref{fig_varden} b shows the evolution of errors with respect to cycles for $p\leq 0.05$ and $\rho\geq 0.80$. These errors are calculated from microsimulations. We can see that approximate errors are able to match simulation values even at high $\rho\geq 0.80$ and low probe proportion $p\leq0.05$ values. When we consider the error differences with respect the mean QLs, they are very close.

%\squeezeup
%\vspace{-10mm}
\section{Estimation with Range Sensor, Unknown $\lambda$, and $p$}
\label{sctest}
%\vspace{-2mm}
Known arrival $\lambda$ and market penetration rate $p$ parameters in Eqs.~(\ref{eqn_ewoq2}) and (\ref{eqn_ewq}) can be relaxed by replacing them with estimators for these low level parameters. Detailed derivations of such estimators were given in (\cite{comert2016queue}). Best performing combinations were adopted from this earlier study. Without Q, the QL with unknown arrival rate and probe proportion can simply be estimated using Eq.~(\ref{eqn_qle}). For $E(N|l,t,m,Q=0$, referred as $Estimator$ $1$ with combination of $\hat{p}_{1}=m/l$ and $\hat{\lambda}_{1}=l/R$ simply Eq.~(\ref{eqn_qle1}) is derived. $Estimator$ $2$ uses $\hat{p}_{2}=\frac{mt}{(mt+(l-m)R)}$ and $\hat{\lambda}_{2}=\frac{l-m}{t}+\frac{m}{R}$ which simplifies to Eq.~(\ref{eqn_qle2}).
\begin{equation}
E(N|l,t,m,Q=0)=l+(1-\hat{p})\hat{\lambda} (R-t)
\label{eqn_qle}
\end{equation}
\begin{equation}
E(N|l,t,m,Q=0)=l+(l-m)(1-t/R)
\label{eqn_qle1}
\end{equation}
\begin{equation}
E(N|l,t,m,Q=0)=l+(1-\frac{mt}{mt+(l-m)R})((l-m)/t+m/R)(R-t)
\label{eqn_qle2}
\end{equation}
%\squeezeup

Similarly, estimation of queues real-time and steady-state including Q are also presented. In this case, QLs can be considered as sum of random variables given in Eq.~(\ref{eqn_ewqi}). Indicator function $I(l\in Q)$ represents the case if the last CV is in the overflow queue, $I(l\in A)$ indicates the last CV in new red arrivals, and $I(l=0)$ indicates that there is no CV in the queue. At any given cycle, these terms represent disjoint events, hence, only one term will be positive where others will be zeros.
\begin{equation}   
E(N_{i}|L=l,T=t,Q_{i}\geq 0)=\begin{cases} I(l\in Q)[l+\hat{\theta} (C-t')+\hat{\theta} R]+\\ I(l\in A)[l+\hat{\theta} \delta]+\\I(l=0)[(1-\hat{p})(E(Q_{i})+\hat{\theta} R)]\end{cases}   
\label{eqn_ewqi}  
\end{equation}
\begin{equation}   
E(N_i|L=l,L=lv,T=t,Q_{i}\geq 0)=\begin{cases} I(l\in Q \land L=lv)[l+\hat{\theta} R]+\\ I(l\in Q \land L \neq lv)[l+1+\hat{\theta} (C-\tau')+\hat{\theta} R]+\\ I(l\in A \land L=lv)[l]+\\ I(l\in A \land L \neq lv)[l+1+\hat{\theta} (R-t')]+\\I(l=0)[(1-\hat{p})(E(Q_i)+\hat{\theta} R)].\end{cases}   
\label{eqn_ewqs}
\end{equation} 

Overflow queue $E(Q_{i})=\frac{Xi(\hat{\rho}-1)}{4} \sqrt{(\hat{\rho}-1)^2+\frac{12(\hat{\rho}-\rho_{o})}{Xi}}$ is adopted from (\cite{Akcelik1980}) where $\rho_{o}$=$0.67+X/600$, $X=24$ vehicles per cycle, $\hat{\rho}=\hat{\lambda} C/X$, and $i=1,2,3...$ denotes the cycle index. For $E(Q_{i})=E(Q)(1-e^{-\beta i})$ from~(\cite{Viti2006}) can also be used which gives very close results (\cite{comert2013simple}) where $E(Q)=\frac{3(\hat{\rho}-\rho_{o})}{2(1-\hat{\rho)}}$. 
%\squeezeup
\begin{figure}[!h]%[!htb]
\centering
\includegraphics[scale=.80]{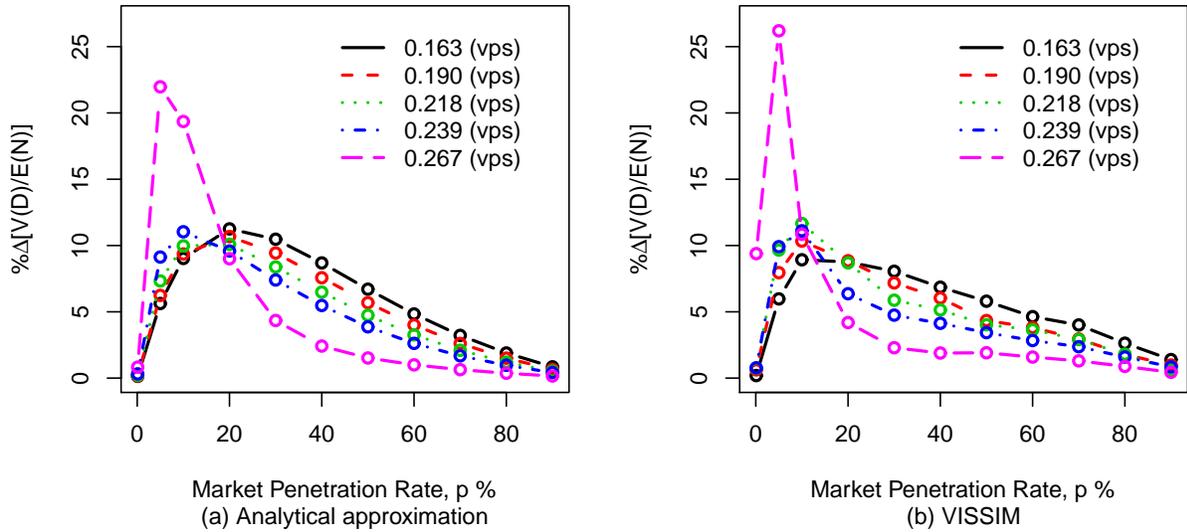}
\caption{Estimation errors $\% \Delta V(D)/E(N)$ with and without range sensor results for different $\lambda$ $vps$ and $ps$}
\label{fig_diffs}       
\end{figure}
%Cycle-to-cycle error of the estimator in Eq.~(\ref{eqn_ewqi}) i.e., $V(D_{i})$ can be given as in Eq.~(\ref{eqn_vdwqi}) with $V(Q_{i})$=$[E(Q)(\hat{\rho}+(1-\hat{p})/0.15)+(\sqrt{\hat{\rho} Xi}-\sigma_{Q_{e}})e^{-\beta i}]^2$ from~\cite{Viti2006}. Where $\sigma_{Q_{e}}$ can be calculated from $\sigma_{Q_{e}}=E(Q)(\hat{\rho}+(1-\hat{\rho})/0.15)$. 
%\begin{equation}   
%V(D_{i}|Q_{i}\geq 0)=\begin{cases} P(l\in Q)[\hat{\theta} %(C-E(T'))+\hat{\theta} R]+\\ P(l\in A)[(1-\hat{p})(1-e^{-p\hat{\lambda} %R})/\hat{p}]+\\P(l=0)[(1-\hat{p})(V(Q_{i})+\hat{\theta} R)]\end{cases}   
%\label{eqn_vdwqi}  
%\end{equation}

\begin{figure}[!h]%[!htb]
\centering
\includegraphics[scale=.80]{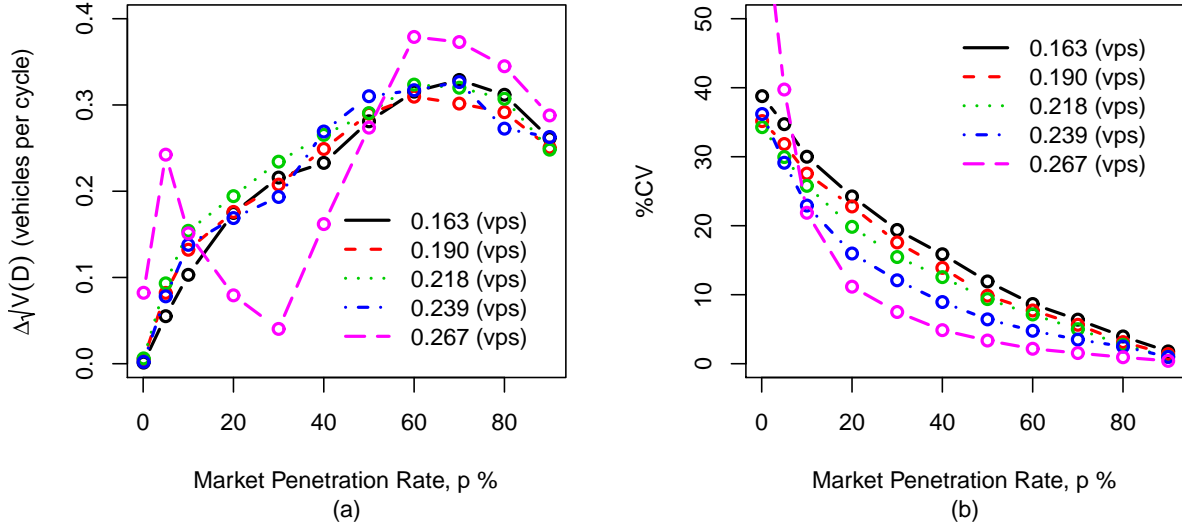}
\caption{Relative estimation errors $\% \Delta \sqrt{V(D)}$ and $\% \sqrt{V(D)}/E(N)$ from VISSIM simulations}
\label{fig_cvs}       
\end{figure}

Average true and estimated QLs derived from VISSIM runs. Note that QLs depend on VISSIM queue definition where default \textit{in queue}=$I(speed\leq10$ kilometers per hour. From Fig.~\ref{fig_cvs}, at 600 vehicles per hour, $7.34$ vehicles per 45 seconds red duration is expected. Similarly others $8.57, 9.79, 10.78$, and $12.01$ vehicles per red. However, overflow queue becomes more important when $\rho>0.80$ on average $0.68$, $2.13$, and $20.68$ vehicles in length for $\rho=0.80$, $\rho=0.88$, and $\rho=0.98$, respectively. Impact of overflow queue also observed accuracy of the estimators start to decrease after $\rho>0.88$ from $\pm$ $1.5$ to $\pm$ $3.45$ vehicles. Overall, $Estimator$ $2$ can provide accuracy between $12\%$ to $19\%$ of the true average QLs. 
%\squeezeup

With overflow queue combination of parameter estimators were input where results are shown in Fig.~\ref{fig_varden}. As queue length estimators closely follow true queue lengths by $p$=$30\%$, queue length estimators with $\lambda$ and $p$ estimators also gets very close to the behavior of estimator with known parameters. 
%\squeezeup
%\vspace{-30pt}
\begin{figure}[!h]%[!htb]
\centering
\includegraphics[scale=.80]{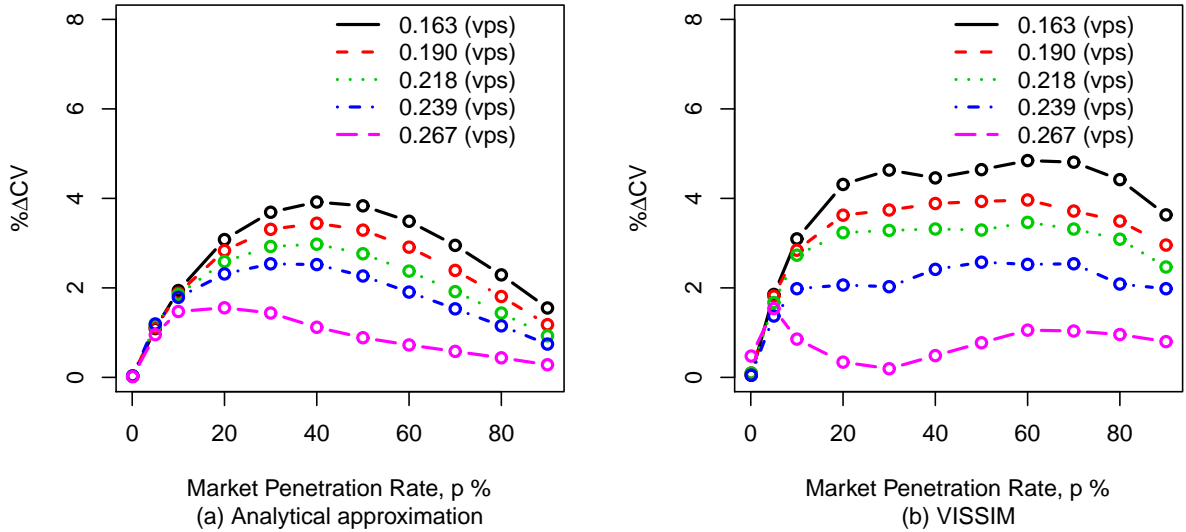}
\caption{Estimation errors $\% \Delta \sqrt{V(D)}/E(N)$ from VISSIM and approximations}
\label{fig_cvup}       
\end{figure}
\vspace{-10pt}
\begin{eqnarray}
E(N_i|l,t,m,Q_{i}=0)=l+(l-m/l)(l/R)(R-t)=l+(l-m)(1-t/R)
\label{eqn_eltapprox2a} 
\end{eqnarray}
\begin{eqnarray}
E(N_i|l,t,m,Q_{i}=0)=l+(1-\frac{mt}{mt+(l-m)R})((l-m)/t+m/R)(R-t)=m+\frac{R(l-m)}{t}
\label{eqn_eltapprox9a} 
\end{eqnarray}
%\vspace{-20pt}

Figs.~\ref{fig_diffs}-\ref{fig_cvup} summarize the QLE errors with the overflow queue. The errors are expressed in $\%\Delta{V(D)}$ and $\%\Delta{CV}$ where, $CV$=$\sqrt{V(D)}/E(N)$. Since estimators are able to capture overflow queue, errors are declining to zero as connected vehicle MPR increases. Results from the literature cited above also confirmed that $p\geq30\%$ queue length can be predicted within $\pm10\%$ across all all $\rho levels$. Similarly at $p=50\%$, queue length can be predicted within $\pm2\%$ in $\%\Delta{CV}$ and $\%\Delta{EN}$.
%\vspace{-30pt}

\begin{figure}[!h]%[!htb]
\centering
\includegraphics[scale=.80]{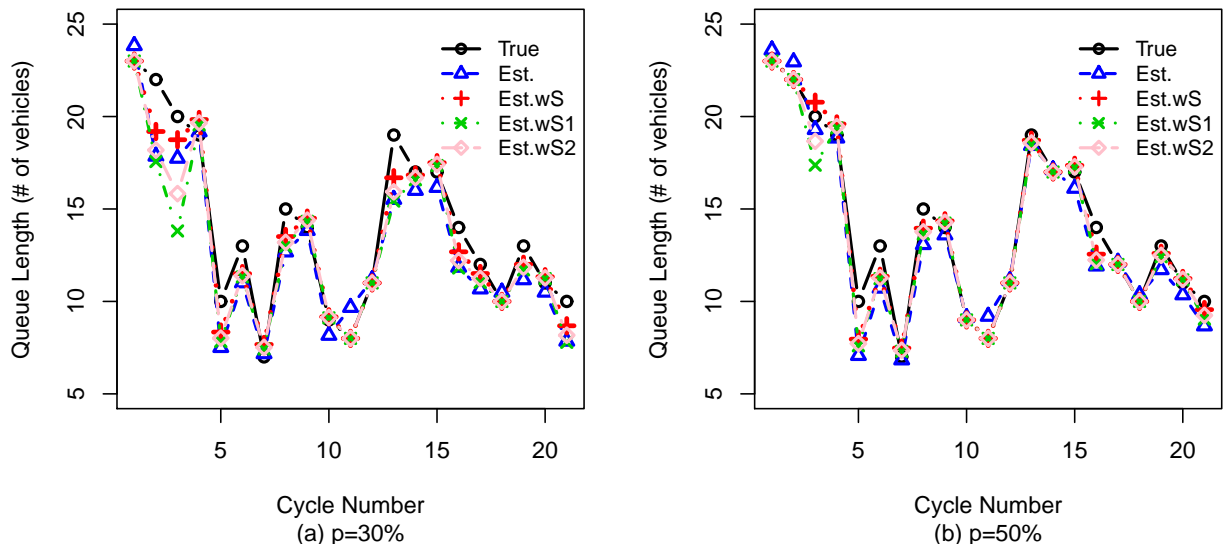}
\caption{Estimation results for $\lambda=0.239$ $vps$ at $p=\{30\%,50\%\}$}
\label{fig_pred30}       
\end{figure}

Fig.~\ref{fig_pred30} confirms these results with good agreement with the true QLs. The figures depict about $20$ cycles for illustration of performance of estimators. Based on the numerical results, for $p\geq30\%$ QLEs with unknown estimators shows good agreement with true average QLs. Thus, any of the $Estimator$ $1$ or $Estimator$ $2$ can be used in accurate queue length estimation in signal control. Estimator with range sensor can also be seen that more closely following the true queue values. 
%\squeezeup
%\vspace{-10mm}

\section{Conclusions}
In this study, models for queue length estimation from connected vehicles with range sensors at signalized intersections are derived. Fully analytical models without overflow queue and approximations with the overflow queue are presented. Contribution of the proposed models are simple input-output models without requiring true queue length observations, as a step towards generalizing estimation problem with partially observed entities in a cycling queueing system, and leverage of simple variations to previously derived models to express additional complexity. 

When range sensor included, estimators are more accurate as high as $25\%$ in estimation errors relative to average queue lengths for high volume-to-capacity ratio $\rho$=$0.98$ at market penetration ratio of $p=5\%$. About $10\%$ improvement is observed when volume-to-capacity ratio is $\leq0.88$. Moreover, approximately $0.25$ vehicle per cycle was improved. For example, at constant arrival rate in $10$ hours, total estimation error of $100$ vehicles (i.e., energy consumption, emissions, and delay) would be improved.
\section*{Acknowledgments}
This study is supported by the Center for Connected Multimodal Mobility ($C^2$$M^2$) (USDOT Tier 1 University Transportation Center) Grant headquartered at Clemson University, Clemson, South Carolina, USA. The authors would also like to acknowledge U.S. Department of Homeland Security (DHS) Summer Research Team Program Follow-On, FY19 US Department of Education MSEIP Grant Award P120A190061, and National Science Foundation (NSF, No. 1719501, 1436222, and 1400991) grants. Any opinions, findings, conclusions or recommendations expressed in this study are those of the author(s) and do not necessarily reflect the views of ($C^2$$M^2$), USDOT, DHS, U.S. Department of Education, or NSF and the U.S. Government assumes no liability for the contents or use thereof. 
\bibliographystyle{elsarticle-harv}
\bibliography{projects}

%\begin{figure}[ht!]
%\centering
%\includegraphics[scale=.25]{allwqrev2}
%\caption{Comparison of estimation errors calculated from analytical formula in %Eq.~(\ref{eqn_vdwq}), approximation in Eq.~(\ref{eqn_vdwqapprox}), and from VISSIM simulations}
%\label{fig_allwq}       
%\end{figure}
%\begin{figure}[ht!]
%\centering
%\includegraphics[scale=.25]{vissimvsrev3}
%\caption{Regression line and \%CV differences of V(D)s calculated from analytical formulas in Eq.~(\ref{eqn_vdwq}), approximation in Eq.~(\ref{eqn_vdwqapprox}), and from VISSIM simulations}
%\label{fig_vissimvs}       
%\end{figure}

%\begin{figure}[ht!]
%\centering
%\includegraphics[scale=.30]{vissimrev2}
%\caption{Comparison of the three cycle-to-cycle estimation errors calculated from %Eq.~(\ref{eqn_vdwqi}) with \cite{Akcelik1980}, Eq.~(\ref{eqn_vdwqi}) with \cite{Viti2006}, and %approximate $V(D_i)$s}
%\label{fig_vissim}       
%\end{figure}

%\begin{figure}[ht!]
%\centering
%\includegraphics[scale=.28]{envsrev4}
%\caption{Expected total queue lengths from VISSIM, approximation, and point queue at different %$\rho$ levels and comparison of the cycle-to-cycle errors from Eq.~(\ref{eqn_vdwqi}) %with~\cite{Viti2006} and approximate $V(D_i)$s}
%\label{fig_en}       
%\end{figure}
\end{document}